\RequirePackage[T1]{fontenc}
\documentclass[conference]{IEEEtran}
\IEEEoverridecommandlockouts
\usepackage{amsmath,amssymb,amsthm}
\usepackage{times}
\usepackage{booktabs}
\usepackage{graphicx}
\usepackage{microtype}
\usepackage{url}
\usepackage{xcolor}
\usepackage{tikz}
\usetikzlibrary{positioning,arrows.meta}

\newtheorem{theorem}{Theorem}
\newtheorem{proposition}{Proposition}

\theoremstyle{definition}

\newcommand{\cI}{\mathcal{I}}

\newcommand{\arxivversionnote}{Author manuscript accepted for publication in the Privacy and Security of Big Data special session at IEEE BigData 2026. \textcopyright{} 2026 IEEE. Personal use of this material is permitted. Permission from IEEE must be obtained for all other uses, in any current or future media, including reprinting/republishing this material for advertising or promotional purposes, creating new collective works, for resale or redistribution to servers or lists, or reuse of any copyrighted component of this work in other works.}

\title{What Unlinked Aggregates Still Prove:\\ Ranker Comparisons from Exact and Differentially Private Item Counts\thanks{\arxivversionnote}}

\author{\IEEEauthorblockN{Chang Xu}
\IEEEauthorblockA{Duke University\\
chang.xu@alumni.duke.edu}
\and
\IEEEauthorblockN{Hao Wang}
\IEEEauthorblockA{Stanford University\\
wanghao@alumni.stanford.edu}
\and
\IEEEauthorblockN{Gedi Zhou}
\IEEEauthorblockA{Independent Researcher\\
gzhou4@alumni.jh.edu}}

\begin{document}
\maketitle

\begin{abstract}
Privacy controls can prevent an evaluator from linking purchases to the recommendations users received, while a merchant can still release complete item-level purchase counts for the evaluation cohort. We ask when these counts identify which recommender performs better. The counts constrain who could have bought each item. We bound the performance difference over all assignments consistent with the counts. A reduced transportation network handles full catalogs and can certify attainable endpoints. If the interval excludes zero, it identifies the better recommender for that cohort. With differentially private counts, one confidence guarantee applies jointly to all comparisons from the same release. In two e-commerce cohorts of about 20,000 and 100,000 users, exact counts narrowed every interval compared with bounds that ignored the counts. Most decisions involved a recommender that favors popular items. For NDCG@10, which rewards ranking purchased items higher, each setting combined one of three personalized pairs with a pattern and rate of missing links. Exact counts settled 14 of 168 settings in the smaller cohort and none of 81 in the larger cohort. In the smaller cohort, exact counts broken down by user group settled two personalized pairs when 90\% of users remained linked, at the cost of revealing more about individual purchases. Whole-cohort private releases with one count per item retained much of the narrowing but settled no personalized pair in either cohort. Grouping less popular items into bins let private counts settle one personalized pair in the smaller cohort.
\end{abstract}


\section{Introduction}

Consider a merchant that commissions an offline comparison of two recommendation systems. The evaluator holds the recommendation list (slate) each system would show to every user, but privacy controls prevent it from joining some purchases to those lists. The merchant still records its own sales. It can therefore count how many users in the evaluation cohort bought each item and return those counts without receiving the slates. The question is whether these unlinked aggregates suffice to compare the two systems on the full cohort.

This separation matters because permission to link data can be selective~\cite{sakshaug2016consent}, and opt-in rules shrink the linkable population~\cite{aridor2023privacy}. We model a first-party merchant that returns complete counts for the cohort and designated purchase (Section~\ref{sec:setting}). Privacy-preserving measurement interfaces offer related release designs: the W3C Attribution draft returns histograms perturbed with differentially private noise~\cite{w3c2026attribution}, and aggregate summary reports are studied by Aksu et al.~\cite{aksu2024summary} and Delaney et al.~\cite{delaney2024dp}. Our evaluator receives the fully known slates, the outcomes of users who remain linked, and item counts for the rest. It seeks a comparison that holds regardless of why users became unlinked.

Common evaluation practice does not use this information. Complete-case evaluation drops unlinked users and changes the cohort. When linkage correlates with outcomes, its conclusion can reverse. Propensity weighting~\cite{schnabel2016recommendations} relies on a model of who stays linked, and imputation~\cite{wang2019doubly} on a model of the outcomes. Worst-case bounds that treat each unlinked user separately need neither model~\cite{manski1989anatomy}, but they ignore the aggregate entirely and are often too wide to support any decision.

The aggregate is informative because it couples users. If exactly three unlinked users bought item $j$, at most three unlinked users whose histories allow $j$ can be credited with it, so a valid bound cannot credit $j$ to every user for whom it is most favorable. Once this coupling is imposed, the sharp bounds on the metric difference are the optimal values of a transportation problem over users and items.

Our contributions are as follows.
\begin{itemize}
\item We formulate ranker comparison from item counts, including public-segment releases. A totally unimodular LP gives sharp paired NDCG@$k$ or Recall@$k$ bounds, nested inside the per-user interval. Refining nested exact-count segments never widens the interval (Section~\ref{sec:method}).
\item A reduced network avoids the user--catalog product, with completion checks and lazy refinement to certify sharpness. For Laplace counts, a simultaneous confidence box covers every comparison from one release with probability at least $1-\alpha$, enabling reuse for future ranker pairs.
\item On Beibei (20,623 users) and Tmall (103,507 users), exact cohort-level counts narrowed every ranker-pair interval under random, trait-dependent, and outcome-dependent linkage. Most decisions involved Pop, whose slates concentrate on few items. Exact segment counts established both tested Beibei personalized pairs at 90\% linkage. Requiring one outcome per user tightened bounds for private and fine-segment releases, while omitting that requirement made little difference with exact cohort-level counts.
\end{itemize}

\section{Setting and a Two-User Example}
\label{sec:setting}

\subsection{A two-user example}

Consider two users whose outcomes cannot be linked and a top-1 comparison. Ranker $A$ recommends item $a$ to both users. Ranker $B$ recommends $b$ to user 1 and $c$ to user 2. Under Hit@1, a user who chose $a$ adds $+1$ to the difference $A-B$, a user who chose the item shown by $B$ adds $-1$, and any other choice adds $0$. Treated separately, each user can contribute anything in $[-1,1]$, so the sum of the two contributions lies in $[-2,2]$ and no comparison is possible.

\begin{figure}[t]
\centering
\begin{tikzpicture}[
  u/.style={draw,rounded corners=2pt,minimum width=12mm,minimum height=5mm,font=\small},
  it/.style={draw,circle,minimum size=6mm,inner sep=0pt,font=\small},
  lab/.style={font=\scriptsize,fill=white,inner sep=1pt},
  >=Stealth]
\node[u] (u1) at (0,0.6) {user 1};
\node[u] (u2) at (0,-0.6) {user 2};
\node[it] (a) at (3.6,1.0) {$a$};
\node[it] (b) at (3.6,0) {$b$};
\node[it] (c) at (3.6,-1.0) {$c$};
\node[font=\scriptsize,right=1mm of a] {count 1};
\node[font=\scriptsize,right=1mm of b] {count 1};
\node[font=\scriptsize,right=1mm of c] {count 0};
\draw[->] (u1) -- node[lab,pos=0.35,above]{$+1$} (a);
\draw[->] (u1) -- node[lab,pos=0.3]{$-1$} (b);
\draw[->] (u2) -- node[lab,pos=0.72,right]{$+1$} (a);
\draw[->] (u2) -- node[lab,pos=0.25]{$0$} (b);
\draw[->,gray] (u2) -- node[lab,pos=0.55]{$-1$} (c);
\end{tikzpicture}
\caption{Item counts couple unlinked users. Arc labels are contributions to Hit@1 of $A$ minus $B$. Each user takes exactly one arc and each item receives exactly its released count, so the two feasible assignments give $+1$ or $0$. Separate per-user bounds would give $[-2,2]$.}
\label{fig:example}
\end{figure}
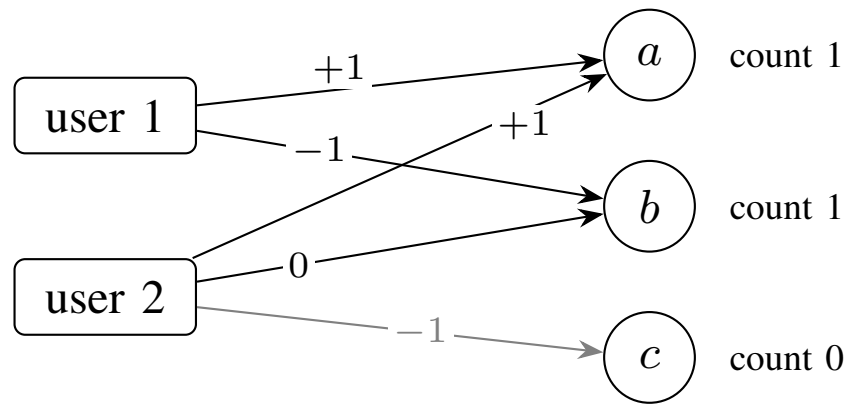

Suppose now that the item-level counts for these two users are released: one unit of $a$ and one unit of an unrelated item $d$. Whichever user chose $a$ contributes $+1$ and the other contributes $0$, because $d$ appears on neither slate. Every assignment consistent with the counts gives a sum of exactly $+1$, so $A$ is proven better on this pair of users even though neither outcome is known. If the counts had instead been one unit of $a$ and one of $b$, the two assignments give $+1$ (user 1 chose $a$) or $0$ (user 1 chose $b$ and user 2 chose $a$), and the sharp interval for the sum is $[0,1]$. Since $0$ is attainable, $A$ is not proven better. Fig.~\ref{fig:example} illustrates this second case as a bipartite assignment, and the general procedure optimizes over all such assignments.

\subsection{Formal setting}

A cohort of users $i=1,\ldots,n$ is evaluated on a catalog $\cI$ of $m$ items. User $i$ has a logged history $H_i\subset\cI$, and its candidates are $C_i=\cI\setminus H_i$. Two fixed rankers produce duplicate-free length-$k$ slates $A_i$ and $B_i$ from $C_i$. Following the leave-one-out protocol that is standard for implicit feedback, each user has one held-out relevant item $s_i\in C_i$~\cite{he2017neural}. For rank $p$, let $w(p)=1/\log_2(p+1)$ for NDCG@$k$ and $w(p)=1$ for Recall@$k$ (which equals Hit@$k$ with one relevant item). Write $w_A(i,j)=w(p)$ if $j$ is at rank $p$ of $A_i$ and $0$ otherwise, define $w_B$ analogously, and let
\begin{equation}
q_i(j)=w_A(i,j)-w_B(i,j),\qquad D=\frac1n\sum_{i=1}^n q_i(s_i).
\label{eq:estimand}
\end{equation}
With one relevant item the ideal DCG is one, so $D$ is exactly the difference in mean NDCG@$k$ (or Recall@$k$) between the two rankers on the full cohort. We call the items with $q_i(j)\neq0$, all of which lie on one of the two slates, the \emph{active} items of user $i$.

\textbf{Linkage.} The evaluator observes $s_i$ for linked users $i\in V$ and does not observe it for unlinked users $i\in W$. The set $W$ may depend on anything, including the outcomes. No missingness model is assumed.

\textbf{Aggregate release.} A public segment label $\sigma(i)\in\{1,\ldots,G\}$ is attached to each user. It is a function of logged information only. Let $W_g=\{i\in W:\sigma(i)=g\}$. The release reports, for each segment $g$ and item $j$, either the exact count
\begin{equation}
c_{gj}=\bigl|\{i\in W_g:\ s_i=j\}\bigr|
\label{eq:counts}
\end{equation}
or integer bounds $\ell_{gj}\le c_{gj}\le h_{gj}$. With $G=1$ this is a single item histogram for the unlinked cohort. Exact counts model first-party counts of the designated outcome over the cohort, from which the known linked outcomes are subtracted. Bounds model a differentially private release (Section~\ref{sec:dp}).

\textbf{Estimand and trust model.} $D$ compares these $n$ users, with no sampling uncertainty. Population conclusions require a separate sampling argument. Replace-one adjacency protects which item a member chose, not cohort membership (Section~\ref{sec:dp}). The cohort, histories, segment labels and sizes, linked set, and linked outcomes are public. Differential privacy protects the remaining outcomes given this information. Reuse concerns one release on this cohort. Releases for further cohorts or time windows compose, and differences between exact releases for overlapping cohorts could expose individual outcomes, so the exact variant assumes one cohort query.

\textbf{Release contract.} The first-party merchant holds purchases. The evaluator sends cohort identifiers, segment labels and the rule selecting one designated test-period purchase per user. The merchant matches purchases internally and returns whole-cohort item counts, exactly or with Laplace noise, without needing linkage status. Subtracting known linked counts gives~\eqref{eq:counts} or the noisy bounds in Section~\ref{sec:dp}. The merchant never receives slates: the privacy control forbids joining purchases to recommendations, while allowing its own sales records. Hiding the internal matching would require a separate secure-computation design, for example using private intersection-sum protocols~\cite{ion2020deploying}. Outside buyers and missing reports yield one-sided bounds (Section~\ref{sec:robust}).

A \emph{completion} assigns an item $s_i\in C_i$ to every unlinked user so that the released counts or bounds hold. The identified set for $D$ is the set of values of~\eqref{eq:estimand} over all completions, and we seek its minimum $L$ and maximum $U$. A comparison is \emph{provable} when $L>0$ (then $A$ is better on the full cohort) or $U<0$ (then $B$ is better).

\section{Sharp Bounds from Unlinked Aggregates}
\label{sec:method}

The section gives guarantees of three kinds. With exact counts, the linear program of Theorem~\ref{thm:sharp} gives sharp bounds: each endpoint is attained by some assignment consistent with the counts. The reduced network of Proposition~\ref{prop:reduced} always gives valid outer bounds, and a completion check establishes when they are also sharp. With noisy counts, the bounds hold with probability at least $1-\alpha$ through a simultaneous confidence box for the released counts (Proposition~\ref{prop:dp}).

\subsection{Baselines without the aggregate}

The generic bound assigns every unlinked user the full metric range, $q_i(s_i)\in[-1,1]$. The per-user bound uses the slates. Because $|C_i|>2k$ in a full catalog, every user has candidates outside both slates, so
\begin{equation}
\underline q_i=\min\Bigl(0,\min_{j\in C_i}q_i(j)\Bigr),\qquad
\overline q_i=\max\Bigl(0,\max_{j\in C_i}q_i(j)\Bigr),
\label{eq:peruser}
\end{equation}
and the cohort interval is
\begin{equation}
\Bigl[\tfrac1n\bigl(\textstyle\sum_{V}q_i(s_i)+\sum_{W}\underline q_i\bigr),\ \tfrac1n\bigl(\sum_{V}q_i(s_i)+\sum_{W}\overline q_i\bigr)\Bigr].
\label{eq:perusercohort}
\end{equation} This is the signed, cardinality-one case of maximized effectiveness difference~\cite{tan2015med}, and it is sharp when unlinked users are unconstrained by any aggregate. Complete-case evaluation reports $\frac{1}{|V|}\sum_{V}q_i(s_i)$, possibly with a confidence interval that is valid when linkage is random. We use it as the comparison point in Section~\ref{sec:results}.

\subsection{Sharp bounds as a transportation problem}

Introduce $x_{ij}\in\{0,1\}$ for $i\in W$ and $j\in C_i$, with $x_{ij}=1$ meaning $s_i=j$. Completions are exactly the integer points of
\begin{equation}
\begin{aligned}
P=\Bigl\{x\ge0:\ &\textstyle\sum_{j\in C_i}x_{ij}=1\ \ (i\in W),\\
&\ell_{gj}\le\textstyle\sum_{i\in W_g}x_{ij}\le h_{gj}\ \ (g,\,j)\Bigr\},
\end{aligned}
\label{eq:polytope}
\end{equation}
with $\ell=h=c$ for exact counts, and the unlinked part of $nD$ is the linear function $\sum_{i,j}q_i(j)x_{ij}$.

\begin{theorem}[Sharp aggregate interval]
\label{thm:sharp}
If $P$ contains an integer point, then
\[
L=\frac1n\Bigl(\sum_{i\in V}q_i(s_i)+\min_{x\in P}\sum_{i,j}q_i(j)x_{ij}\Bigr)
\]
and $U$ (with $\max$ in place of $\min$) are the minimum and maximum of $D$ over all completions, and both are attained by completions.
\end{theorem}
\begin{proof}
Every variable $x_{ij}$ appears with coefficient one in exactly one user row and in exactly one segment-item row, so the constraint matrix is the incidence matrix of a bipartite graph and is totally unimodular. Appending its negation and the identity preserves this property. Since all right-hand sides are integers, every vertex of $P$ is integral~\cite{hoffman1956integral,schrijver2003combinatorial}. An integral point of $P$ has one unit per user and satisfies the counts, so it is a completion. Conversely every completion lies in $P$. A linear objective over a nonempty bounded polytope is optimized at a vertex, so the LP optimum is attained by a completion, and no completion can exceed it.
\end{proof}

\begin{proposition}[Information ladder]
\label{prop:ladder}
Let the generic, per-user, and aggregate intervals be computed from the same linked outcomes. Then generic $\supseteq$ per-user $\supseteq$ aggregate. If a segmentation $\sigma'$ refines $\sigma$ and the exact counts are released for $\sigma'$, the aggregate interval under $\sigma'$ is contained in that under $\sigma$. With singleton segments the interval collapses to the point $D$.
\end{proposition}
\begin{proof}
Each step shrinks a feasible set: the generic range relaxes the slate structure, per-user completions ignore the counts, counts for $\sigma'$ imply the counts for $\sigma$ by summation, and singleton counts reveal every $s_i$. Minimum and maximum over a subset are nested.
\end{proof}

Proposition~\ref{prop:ladder} gives the utility side of the trade-off studied in Section~\ref{sec:q3}: finer segments can prove more, but a segment with few unlinked users also reveals more about each of them.

\subsection{A reduced network that scales with the cohort}

A direct formulation of~\eqref{eq:polytope} has up to $|W|\cdot m$ variables, which is too large for big catalogs. Most of these variables carry zero contribution. We therefore solve a reduced problem. User $i$ either chooses one of its active items $j$ with $q_i(j)\ne0$ at cost $q_i(j)$, or sends its unit to a zero-cost pool node of its segment. The pool distributes units to items, and the segment-item bounds apply to the sum of direct and pooled units. Items that are active for no user in a segment are merged into one pool bucket whose bounds are the sums of their bounds. Any integer total within those sums splits back into admissible item counts.

\begin{proposition}[Reduced network]
\label{prop:reduced}
The reduced problem has $O(|W|k+Gm')$ arcs, where $m'$ is the number of items active for at least one unlinked user, and its optimal values $L_R,U_R$, after adding the linked term and dividing by $n$, satisfy $L_R\le L$ and $U_R\ge U$. If, for some split of the merged bucket into item counts, the pooled units of an optimal reduced solution can be matched to the pooled users so that every user receives an item in $C_i$ that is not active for it, then $L_R=L$ (respectively $U_R=U$).
\end{proposition}
\begin{proof}
Every completion maps to a feasible reduced solution with the same objective, by routing each user whose $s_i$ is not active for it through the pool, so the reduced optima are outer bounds. If the matching exists, the reduced optimum is itself the objective of a completion, because pooled users then receive inactive, admissible items with contribution zero. An outer bound attained by a feasible point is the exact optimum.
\end{proof}

The reduced LP is again a network problem with integral vertices. The matching in Proposition~\ref{prop:reduced} is a bipartite feasibility check between pooled users and pooled units in which each user excludes only its history and at most $2k$ active items. We first try a greedy assignment that serves the most constrained users first, repair it by one-step swaps, and fall back to an exact maximum flow on pooled graphs of at most $5\times10^6$ arcs. Otherwise the endpoint stays uncertified. We call an endpoint \emph{certified} when the check succeeds. For the binned and multi-item variants of Section~\ref{sec:results} we use the greedy pass only. The check tries one split of the merged bucket, so it is sufficient for sharpness but not necessary.

The check fails mainly when an item is active for most users, as happens with popularity slates: pooled units of that item cannot go to users for whom it would carry a contribution. We then refine lazily. Every item that receives pooled units and is active for some pooled user joins a set $S$. Items in $S$ receive explicit arcs from every user whose history admits them (with cost $q_i(j)$, which is zero for inactive users), and the pool may no longer route to them. The refined problem is still a relaxation of~\eqref{eq:polytope} and exact on $S$, so its optimum is again a valid bound. The check is repeated for at most eight rounds (three for the Tmall private releases of Pop pairs) or until $|S|\,|W|$ exceeds $4\times10^6$ arcs. The last bound is then reported as a valid but uncertified outer bound.

\subsection{Differentially private counts}
\label{sec:dp}

When the counts leave a trusted aggregator, they can be protected with the Laplace mechanism~\cite{dwork2006calibrating}. For each released bin $(g,j)$ the aggregator publishes $\hat c_{gj}=c^{\mathrm{all}}_{gj}+Z_{gj}$, where $c^{\mathrm{all}}_{gj}$ counts relevant items of all cohort users in segment $g$ and $Z_{gj}\sim\mathrm{Laplace}(2/\epsilon)$ independently. Under replace-one adjacency, changing one user's relevant item moves one unit between two bins, so the $L_1$ sensitivity is 2 and the release is $\epsilon$-differentially private with respect to any single user's outcome, hence also for each unlinked outcome given the linked ones. For add/remove adjacency, a histogram with at most one outcome per user has sensitivity 1 and uses scale $1/\epsilon$. Protecting cohort membership would also require a design that does not expose membership and segment sizes as public inputs. The scale $2/\epsilon$ for unit contributions is consistent with the budget accounting of the W3C Attribution draft~\cite{w3c2026attribution}.

Let $M$ be the number of released bins ($M=Gm$ when every item is released in every segment), $t=(2/\epsilon)\ln(M/\alpha)$, and $c^V_{gj}$ the known count among linked users. The evaluator uses the box
\begin{equation}
\ell_{gj}=\max\bigl(0,\lceil \hat c_{gj}-t-c^V_{gj}\rceil\bigr),\quad
h_{gj}=\max\bigl(0,\lfloor \hat c_{gj}+t-c^V_{gj}\rfloor\bigr),
\label{eq:box}
\end{equation}
together with the public segment sizes $|W_g|$, which the user rows of~\eqref{eq:polytope} already enforce.

\begin{proposition}[Private certificate]
\label{prop:dp}
With probability at least $1-\alpha$ over the noise, the true unlinked counts lie in the box~\eqref{eq:box} for all bins simultaneously. On this event the interval of Theorem~\ref{thm:sharp} computed from the box contains $D$ for every pair of rankers evaluated on the same cohort. Hence the rule ``$A$ if $L>0$, $B$ if $U<0$, abstain otherwise'' issues a wrong direction for any comparison, or for any collection of comparisons, with probability at most $\alpha$.
\end{proposition}
\begin{proof}
$\Pr(|Z_{gj}|>t)=e^{-t\epsilon/2}=\alpha/M$, and a union bound over the $M$ bins gives simultaneous coverage. On that event the true counts define a completion inside $P$, so its value $D$ lies in $[L,U]$ for any objective, and a wrong sign would require $D$ outside $[L,U]$. The intervals are post-processing of the release and consume no further privacy budget.
\end{proof}

\begin{proposition}[Cost of privacy]
\label{prop:gap}
Suppose every unlinked user admits every item with a nonzero bound (no history exclusions among these items). On the coverage event of Proposition~\ref{prop:dp}, the private interval $[L_\epsilon,U_\epsilon]$ and the exact-count interval $[L,U]$ satisfy, with $t=(2/\epsilon)\ln(M/\alpha)$,
\[
0\le U_\epsilon-U\le \frac{2tM}{n},\qquad 0\le L-L_\epsilon\le \frac{2tM}{n}.
\]
\end{proposition}
\begin{proof}
Nesting holds because the true counts lie in the box. Let $c'$ be the count vector of an assignment attaining $U_\epsilon$. Since $c$ and $c'$ both lie in the box, $\|c'-c\|_1\le 2tM$, and both sum to $|W|$. Reassigning at most $\|c'-c\|_1/2$ users turns this assignment into one with counts $c$, which is feasible under the stated condition, and each reassignment changes the sum of contributions by at most 2 because $q_i(j)\in[-1,1]$. Hence $nU\ge nU_\epsilon-2tM$. The lower endpoint is symmetric.
\end{proof}

The bound is a design guide for the case without history exclusions. With the release design fixed, it shrinks as $1/n$. At fixed $n$, it grows with $M\ln M$, which favors releases with few bins as long as merging items does not widen the exact interval by more than it saves. For the full histograms at $\epsilon=1$ and $\alpha=0.05$, $2tM/n$ is 18.5 on Beibei and 4.7 on Tmall, exceeding the metric-difference range. Its useful regime is therefore a release with sufficiently few bins, such as the tiered designs below. Section~\ref{sec:results} measures the gap directly with history exclusions.

If $P$ is empty for the box, which can only happen outside the coverage event, the release is inconsistent with the segment sizes and the procedure abstains. Clipping at zero and rounding to integers only remove impossible values, so the coverage event is unchanged.

\section{Evaluation Design}
\label{sec:design}

We ask four questions. \emph{Q1:} How much of the uncertainty left by per-user bounds do exact item counts remove, and how often does this turn an abstention into a decision? \emph{Q2:} How much survives when the counts are released with differential privacy, and how does this depend on cohort size? \emph{Q3:} How do segment-level releases trade tighter bounds against inference about individual outcomes? \emph{Q4:} Does one release support many comparisons, and is the reduced network sharp and fast in practice? Section~\ref{sec:robust} adds robustness checks.

\subsection{Data, protocol, and rankers}

\textbf{Beibei.} The Beibei e-commerce data~\cite{gao2019neural}, in the release of~\cite{kim2024beibei}, provide 21,716 users, 7,977 items, and train, validation, and test purchase splits. The cohort consists of all 20,623 users with both training history and a test purchase. For each user one test purchase is designated as the relevant item by a fixed keyed hash. The user's other test purchases, if any, remain ordinary candidates. The history $H_i$ is the union of the user's training and validation items. Models are fit on the 217,391 training pairs.

\textbf{Tmall.} The IJCAI-15 Tmall release~\cite{tianchi2015tmall} contains the purchase, click, cart, and favorite logs of Tmall users from May to November 2014. We use the 3,292,144 purchase events of 424,170 buyers. The catalog is the 10,000 items with the most distinct buyers, and purchases of other items are discarded. A user's last purchase day is the last day with a catalog purchase. The history $H_i$ is the set of catalog items bought on earlier days, and the relevant item is chosen by a keyed hash among the items bought on the last day that are not in $H_i$, the other last-day purchases remaining ordinary candidates. The cohort consists of the 103,507 users with at least three history items and such a relevant item, and models are fit on the histories. The release also records each user's age range, which we use for one linkage mechanism below.

\textbf{Protocol.} Rankings are computed over the full catalog after removing each user's history, without sampled negatives, following the recommendation of Krichene and Rendle~\cite{krichene2020sampled}. Slates have length $k=10$.

\textbf{Rankers.} We use four standard rankers with fixed hyperparameters: item popularity (Pop), item-based nearest neighbors with cosine similarity, 100 neighbors, and shrinkage 10 (ItemKNN)~\cite{sarwar2001item}, PureSVD with 64 factors~\cite{cremonesi2010performance}, and EASE with $\lambda=250$~\cite{steck2019ease}. We compare all six pairs rather than selecting pairs, so both clearly separated comparisons and near ties appear. Table~\ref{tab:data} reports each ranker's quality.

\begin{table}[t]
\caption{Full-catalog quality of the four rankers (one relevant item per user, $k=10$) and the number of distinct items that appear on their slates. Beibei: 20,623 users, 7,977 items. Tmall: 103,507 users, 10,000 items.}
\label{tab:data}
\centering
\footnotesize
\setlength{\tabcolsep}{2.4pt}
\begin{tabular}{lrrrrrr}
\toprule
 & \multicolumn{3}{c}{Beibei} & \multicolumn{3}{c}{Tmall} \\
\cmidrule(lr){2-4}\cmidrule(lr){5-7}
Ranker & NDCG & Recall & Slate items & NDCG & Recall & Slate items \\
\midrule
Pop & .0402 & .0815 & 22 & .0019 & .0050 & 15 \\
ItemKNN & .0576 & .1061 & 6,623 & .0058 & .0110 & 9,307 \\
PureSVD & .0352 & .0658 & 303 & .0044 & .0085 & 592 \\
EASE & .0787 & .1413 & 1,042 & .0065 & .0124 & 7,422 \\
\bottomrule
\end{tabular}
\end{table}

\subsection{Linkage patterns}

We call one combination of ranker pair, metric, linked fraction, and linkage mask a \emph{cell}. On Beibei, for each pair, metric, and linked fraction in $\{0.5,0.7,0.8,0.9\}$ we form fourteen masks, each linking exactly the rounded fraction of users. Ten \emph{hash} masks choose linked users by keyed hashes of user identifiers and are independent of outcomes by construction. Two \emph{activity} masks keep the most active or the least active users linked, by history length. They model linkage that depends on an observable user trait. These deterministic masks leave some history-length strata entirely unlinked, violating the overlap condition needed for propensity weighting. An outcome model must extrapolate into those strata. Two \emph{oracle} masks keep linked the users with the lowest or highest realized contribution $q_i(s_i)$. They are stress tests of the guarantee under arbitrary missingness, and they show how complete-case estimates fail. On Tmall we use NDCG, linked fractions $\{0.5,0.7,0.9\}$, and nine masks: three hash, the two activity, the two oracle, and two \emph{demographic} masks that draw linked users without replacement with probability proportional to $\exp(\pm a_i)$, where $a_i$ is the standardized age range (missing values set to the mean). The full-cohort difference $D$ is used only to check the results.

\subsection{Information levels and outputs}

For every cell we compute the generic interval, the per-user interval~\eqref{eq:perusercohort}, and the sharp interval from exact cohort-level item counts ($G=1$). NDCG intervals are also computed from a Laplace release of the full item histogram ($M=m$ bins) with $\alpha=0.05$ and one noise draw per cell: on the first five Beibei hash masks at $\epsilon\in\{0.5,1,2,4\}$, and on the three Tmall hash masks at $\epsilon\in\{1,4\}$. For Q2 we additionally draw nested hash subsamples of both cohorts and evaluate tiered releases (Section~\ref{sec:results}-B). For Q3 we release counts per segment for $G\in\{1,4,16,64,256\}$ segments, formed either by $k$-means on history embeddings (\emph{profile} segments) or by hashing user identifiers (\emph{random} segments). Neither uses outcomes. On Beibei it uses EASE vs.\ ItemKNN and ItemKNN vs.\ PureSVD, NDCG, linked fractions 0.5, 0.7, and 0.9, and two hash masks. On Tmall, we use random segments for the same two pairs at 90\% linked. For Q4 we draw 50 releases per $\epsilon$ on Beibei (70\% linked, first hash mask) and use each for all six comparisons. LPs are solved with the HiGHS dual simplex~\cite{huangfu2018highs}. We rebuilt the Tmall arrays from the Tianchi files and verified the extracted purchase file against its recorded checksum. Implementation code and instructions for accessing the source data are available at \url{https://github.com/AnthonyXu109/psbd26-artifact}.

\section{Results}
\label{sec:results}

\subsection{Q1: Exact item counts}

\begin{figure}[t]
\centering
\includegraphics[width=\columnwidth]{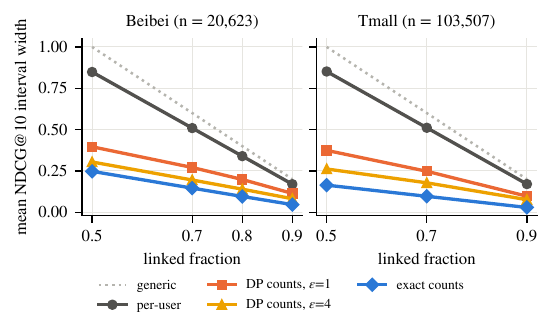}
\caption{Mean NDCG@10 interval width over all six ranker pairs on hash masks (Beibei: the first five of ten hash masks. Tmall: three). Dotted: generic range. One Laplace draw per cell (Section~\ref{sec:q4} repeats draws). Mask ranges are at most 0.007, so the shaded bands nearly coincide with the curves.}
\label{fig:ladder}
\end{figure}

Exact item counts narrowed the interval for every ranker pair, and a narrower interval decides more comparisons. Fig.~\ref{fig:ladder} shows mean NDCG@10 widths on hash masks. The per-user interval remains wide: at 50\% linked it spans 0.85 on Beibei and 0.85 on Tmall. Exact cohort-level item counts shrink it to 0.25 and 0.16. Averaged over pairs, all masks, and linked fractions, the exact-count width is 30\% of the per-user width for NDCG and 38\% for Recall on Beibei, and 20\% for NDCG on Tmall. 

Across 336 Beibei NDCG cells and 162 Tmall cells, exact counts proved a direction in 25\% and 41\%, respectively (Beibei Recall: 22\%). The per-user interval proved no NDCG direction and only one Recall direction. Every exact-count interval contained the full-cohort difference, and no decision was wrong under any tested mask. Table~\ref{tab:decisions} shows that decisions concentrate on Pop: without it, only 14 of 168 Beibei cells and none of 81 Tmall cells were decided. Pop places the same few items on almost every slate (Table~\ref{tab:data}), so their counts tightly constrain its possible credit. In contrast, Tmall's personalized differences of 0.0008 to 0.0021 remain unresolved at the cohort level.

\begin{table}[t]
\caption{Provable NDCG@10 directions from exact item counts: number of masks with a proven direction at each linked fraction (of 14 masks on Beibei and 9 on Tmall). Every proven direction agrees with the sign of the full-cohort difference $D$.}
\label{tab:decisions}
\centering
\footnotesize
\setlength{\tabcolsep}{3.1pt}
\begin{tabular}{llrrrrr}
\toprule
Cohort & Pair ($A$ vs.\ $B$) & $D$ & 50\% & 70\% & 80\% & 90\% \\
\midrule
Beibei & EASE vs.\ ItemKNN & $+0.0211$ & 0 & 0 & 0 & 1 \\
Beibei & EASE vs.\ PureSVD & $+0.0435$ & 0 & 0 & 0 & 13 \\
Beibei & EASE vs.\ Pop & $+0.0385$ & 2 & 13 & 13 & 13 \\
Beibei & ItemKNN vs.\ PureSVD & $+0.0224$ & 0 & 0 & 0 & 0 \\
Beibei & ItemKNN vs.\ Pop & $+0.0174$ & 1 & 2 & 12 & 13 \\
Beibei & PureSVD vs.\ Pop & $-0.0050$ & 0 & 0 & 0 & 0 \\
\midrule
Tmall & EASE vs.\ ItemKNN & $+0.0008$ & 0 & 0 & -- & 0 \\
Tmall & EASE vs.\ PureSVD & $+0.0021$ & 0 & 0 & -- & 0 \\
Tmall & EASE vs.\ Pop & $+0.0046$ & 8 & 8 & -- & 8 \\
Tmall & ItemKNN vs.\ PureSVD & $+0.0014$ & 0 & 0 & -- & 0 \\
Tmall & ItemKNN vs.\ Pop & $+0.0038$ & 8 & 8 & -- & 8 \\
Tmall & PureSVD vs.\ Pop & $+0.0024$ & 2 & 8 & -- & 8 \\
\bottomrule
\end{tabular}
\end{table}

Decisions depend on the slates as well as on the effect size: on Beibei, ItemKNN against PureSVD (0.0224) is never decided while ItemKNN against Pop (0.0174) is.

Complete-case analysis drops unlinked users. On Beibei its mean absolute NDCG error is 0.0007, 0.0036 and 0.0503 under hash, activity and oracle masks. Under random linkage, approximate normal 95\% intervals are valid and more powerful: they covered $D$ in every hash cell, proving a direction in 99\% of Beibei and 100\% of Tmall cells. Coverage under activity masks fell to 51\% and 94\%. Under oracle masks, across cohorts and metrics, they asserted a wrong direction in 66 of 132 cells. Point estimates reversed sign in 48 Beibei cells (44 with $|D|\ge0.005$) and 18 Tmall cells, all under oracle masks. The aggregate procedure abstained in every reversal.

\subsection{Q2: Differentially private counts and cohort size}

On Beibei, a Laplace release of the full 7,977-item histogram keeps a large part of the exact-count gain (Fig.~\ref{fig:ladder}). Retention, the share of the per-user-to-exact width reduction that survives (100\% keeps all of it and 0\% keeps none), averaged over cells, is 46\%, 58\%, 71\%, and 81\% at $\epsilon=0.5$, 1, 2, and 4. On Tmall the 10,000-item release keeps 56\% at $\epsilon=1$ and 74\% at $\epsilon=4$. On both cohorts retention falls as the linked fraction rises, because the noise per bin is fixed while the unlinked counts shrink. On Tmall at $\epsilon=1$ it falls from 64\% at 50\% linked to 46\% at 90\%. Directions were proven in 16\% of the Beibei cells with a private release at $\epsilon=1$ and 18\% at $\epsilon=4$, against 25\% with exact counts on the same cells. On Tmall the corresponding shares are 28\%, 43\%, and 44\%. Table~\ref{tab:privatepairs} reports the decisions by pair on these same hash cells. Every private decision from a full item histogram involved Pop. With those histograms, the five exact-count Beibei decisions between EASE and PureSVD were lost at both privacy levels. The tiered release below settled this pair in two settings. Private intervals contained $D$ in every cell, including draws in which the box missed a count. The simultaneous box covered the true unlinked counts in 97.3\% of Beibei draws and 96.3\% of Tmall draws, consistent with the 95\% target.

\begin{table}[t]
\caption{Proven NDCG@10 directions on the private-release hash grids, by pair. Each entry is a count of decided cells. There are 20 cells per Beibei pair and 9 per Tmall pair. Exact counts use the same cells. All decisions agree with the full-cohort direction.}
\label{tab:privatepairs}
\centering
\footnotesize
\setlength{\tabcolsep}{3.5pt}
\begin{tabular}{llrrr}
\toprule
Cohort & Pair & Exact & $\epsilon=1$ & $\epsilon=4$ \\
\midrule
Beibei & EASE vs.\ ItemKNN & 0 & 0 & 0 \\
 & EASE vs.\ PureSVD & 5 & 0 & 0 \\
 & EASE vs.\ Pop & 15 & 14 & 15 \\
 & ItemKNN vs.\ PureSVD & 0 & 0 & 0 \\
 & ItemKNN vs.\ Pop & 10 & 5 & 7 \\
 & PureSVD vs.\ Pop & 0 & 0 & 0 \\
\midrule
Tmall & EASE vs.\ ItemKNN & 0 & 0 & 0 \\
 & EASE vs.\ PureSVD & 0 & 0 & 0 \\
 & EASE vs.\ Pop & 9 & 6 & 9 \\
 & ItemKNN vs.\ PureSVD & 0 & 0 & 0 \\
 & ItemKNN vs.\ Pop & 9 & 6 & 8 \\
 & PureSVD vs.\ Pop & 6 & 3 & 6 \\
\bottomrule
\end{tabular}
\end{table}

\begin{figure}[t]
\centering
\includegraphics[width=\columnwidth]{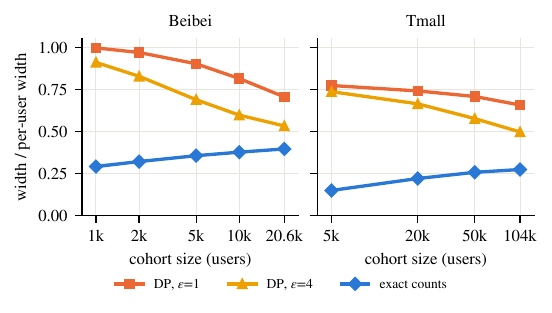}
\caption{Width of the aggregate interval relative to the per-user interval on nested subsamples (NDCG@10, 70\% linked). Beibei: EASE vs.\ ItemKNN and ItemKNN vs.\ PureSVD, five hash masks. Tmall: EASE vs.\ ItemKNN and EASE vs.\ Pop (one undecided and one decided pair), two hash masks. Lower is better. Across masks the ratios differ by at most 0.03 (shaded).}
\label{fig:scale}
\end{figure}

The cost of privacy falls with cohort size in these runs (Fig.~\ref{fig:scale}). On nested Beibei subsamples from 1,000 to 20,623 users, the exact-count width ratio rises from 0.29 to 0.40 as more users can absorb each count and the coupling loosens. Private ratios instead fall from 1.00 to 0.71 at $\epsilon=1$ and from 0.91 to 0.53 at $\epsilon=4$: counts grow while the noise scale stays fixed. Tmall shows the same pattern from 5,000 to 103,507 users, with the private ratio at $\epsilon=1$ falling from 0.77 to 0.66 and the exact ratio rising from 0.15 to 0.27.

\textbf{Tiered item bins.} Full histograms have thousands of zero or one counts below the noise scale, and $t$ grows with the number of bins. We therefore release the $m_0$ historically most popular items individually and merge the rest into eight bins with roughly equal historical popularity. This map is fixed before release and independent of rankers, preserving reuse. Each variable still enters one user row and one bin row, so Theorem~\ref{thm:sharp} and Proposition~\ref{prop:dp} apply with $M=m_0+8$.

Table~\ref{tab:bins} reports all six pairs. Beibei-2k uses a nested 2,000-user subsample, three hash masks and 70\% linkage. The full Beibei grid uses two masks at 50\%, 70\% and 90\%, and Tmall uses two masks at 70\%. To compare retention across these grids, we fix $m_0=100$, their common value, and average each cell's retained gain relative to its full-histogram exact width. At $\epsilon=1$, tiered bins raise retention from 24\% to 49\% on Beibei-2k, from 60\% to 77\% on Beibei and from 58\% to 71\% on Tmall. Merging the tail loses exact information, as the table shows, but the smaller noisy release more than compensates at $\epsilon=1$. At $\epsilon=4$ and 90\% linkage, the $m_0=100$ release also established EASE over PureSVD under both tested Beibei hash masks.

\begin{table}[t]
\caption{Mean NDCG@10 interval width for tiered releases. The $m_0$ most popular items are released individually and the rest in eight bins. The row labeled all denotes a full histogram.}
\label{tab:bins}
\centering
\footnotesize
\setlength{\tabcolsep}{4pt}
\begin{tabular}{lrrrr}
\toprule
Cohort & $m_0$ & Exact & $\epsilon=4$ & $\epsilon=1$ \\
\midrule
Beibei-2k & 10 & 0.210 & 0.222 & 0.257 \\
Beibei-2k & 30 & 0.195 & 0.217 & 0.265 \\
Beibei-2k & 100 & 0.178 & 0.231 & 0.311 \\
Beibei-2k & all & 0.123 & 0.323 & 0.406 \\
Beibei-2k & per-user & \multicolumn{3}{c}{0.509} \\
\midrule
Beibei & 100 & 0.187 & 0.194 & 0.213 \\
Beibei & 1000 & 0.157 & 0.187 & 0.237 \\
Beibei & all & 0.146 & 0.193 & 0.260 \\
Beibei & per-user & \multicolumn{3}{c}{0.509} \\
\midrule
Tmall & 100 & 0.198 & 0.199 & 0.201 \\
Tmall & all & 0.096 & 0.178 & 0.248 \\
Tmall & per-user & \multicolumn{3}{c}{0.511} \\
\bottomrule
\end{tabular}
\end{table}

\subsection{Q3: Segment-level releases and disclosure}
\label{sec:q3}

\begin{figure}[t]
\centering
\includegraphics[width=\columnwidth]{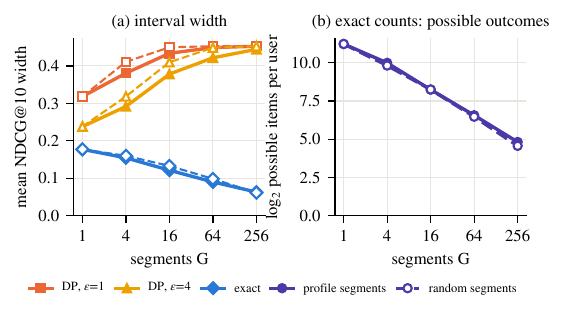}
\caption{Segment-level releases on Beibei (NDCG@10, 70\% linked, two pairs, two hash masks). (a) Interval width (first-pass outer bounds). Solid lines use profile segments and dashed lines random segments. (b) Mean $\log_2$ of the exact number of items an unlinked user could have chosen given the exact segment counts and all histories. Lower values mean more disclosure.}
\label{fig:segments}
\end{figure}

Finer segments tighten the exact-count interval, widen the private one, and reveal more about individual outcomes (Fig.~\ref{fig:segments}). With exact counts the interval tightens as segments are refined: at 70\% linked the mean width falls from 0.177 with one segment to 0.063 with 256 profile segments. Random segments hash users modulo $G$ with each $G$ dividing the next, so they are nested and Proposition~\ref{prop:ladder} applies. The $k$-means profile segments are fitted separately for each $G$. For both types the first-pass outer bounds (69\% certified sharp) decrease with $G$ in every cell. At 90\% linked, exact segment counts decide EASE against ItemKNN in both masks from $G=4$ onward and ItemKNN against PureSVD from $G=16$ onward, while neither is decided with $G=1$ in these two masks. With $G\le16$ no unlinked outcome is pinned to a single item. On Tmall at 90\% linked, random segment counts narrow EASE against ItemKNN from 0.0356 with one segment to 0.0026 with 256 segments of about 40 unlinked users and ItemKNN against PureSVD from 0.0406 to 0.0041 (two hash masks). No interval excludes zero, and the EASE lower endpoint lies within 0.0002 of it.

Panel (b) measures disclosure through each unlinked user's feasible items, those assigned to it by some completion consistent with the counts and histories. We find them from a true reference assignment and alternating cycles in its residual graph, using one strongly-connected-component pass per segment. An outcome is \emph{determined} if only one item is feasible. The mean $\log_2$ size of this set falls from 11.2 bits with one segment to 4.8 bits with 256 profile segments. At 70\% linked at most 0.02\% of Beibei outcomes are determined by the counts even with 256 segments, and at 90\% linked at most 0.7\%. In the 2,000-user subsample, where 256 random segments hold about two unlinked users each, 11\% of unlinked outcomes at 70\% linked are determined, against none with 64 or fewer segments. Segment size alone does not bound disclosure, since a segment whose members all bought the same item reveals every outcome. A formal guarantee requires the noise of Section~\ref{sec:dp}. These are exact-reconstruction measures. They do not bound partial or approximate reconstruction~\cite{garfinkel2019understanding} or inference that uses ranker scores as a prior.

With private counts, refinement widens the interval instead. Each segment adds a full histogram of noisy bins, so the noise per bin stays fixed while the counts per bin shrink. At $\epsilon=1$ the width grows from 0.318 with one segment to 0.453 with 256, equal to the per-user width, for profile and random segments alike.

\subsection{Robustness checks}
\label{sec:robust}

Beyond the extreme oracle masks, on Beibei we draw linked users without replacement with probability proportional to $\exp(\pm z_i)$, where $z_i$ is the standardized log popularity of the purchased item (NDCG, all pairs, linked fractions 0.5, 0.7, 0.9, five draws per direction). The complete-case 95\% interval covered $D$ in 19\% of these 180 cells and asserted a wrong direction in 15. Its mean absolute error was 0.0106 against a mean $|D|$ of 0.0246. The aggregate intervals contained $D$ everywhere and proved a direction in 36\%. On Tmall the demographic masks tie linkage to the recorded age range in the same way. Here the linked subset remained representative: the complete-case interval covered $D$ in all 36 cells and proved a direction in each, with a mean point error of 0.00017 against a mean $|D|$ of 0.00252. The aggregate intervals contained $D$ in every cell and proved a direction in 47\% of them, close to the share under hash masks.

Counts contaminated by outside-cohort purchases still give one-sided bounds. For totals $T_j=c^{\mathrm{all}}_j+O_j$, where $O_j=\mathrm{round}(\rho c^{\mathrm{all}}_j)$ is unknown, the evaluator uses $\ell_j=0$ and $h_j=T_j-c^V_j$. On Beibei at 70\% linkage, the EASE--ItemKNN width grows from 0.163 with exact counts to 0.314 at $\rho=1$, versus 0.376 without counts. EASE--Pop remains decided in all five hash masks at $\rho=0.1$, but none at $\rho\ge0.5$.

Under-reporting instead leaves unallocated outcomes. Given reported counts $\tilde c_j$ and known $|W|$, use $\ell_j=\tilde c_j$ and $h_j=\tilde c_j+|W|-\sum_{j'}\tilde c_{j'}$. When a random share $\rho'$ is omitted, the EASE--ItemKNN width rises from 0.163 to 0.185 at $\rho'=0.05$ and 0.247 at $\rho'=0.2$. Even 5\% under-reporting removes all five EASE--Pop decisions. Thus complete cohort-restricted reports preserve information that these defects can erase.

\textbf{Several relevant items per user.} Treating all of a Beibei user's test purchases as relevant (25\% of users have more than one), with counts $r_i$ known to the evaluator, changes the user rows of~\eqref{eq:polytope} to $\sum_j x_{ij}=r_i$ with $x_{ij}\le1$, normalizes NDCG by the ideal DCG for $r_i$ relevant items and Recall by $r_i$, and keeps the matrix totally unimodular. On six pairs, five hash masks, and linked fractions 0.5 to 0.9, exact counts reduce the per-user width to 36\% for NDCG and 50\% for Recall (single-item, same cells: 29\% and 38\%) and prove a direction in 22\% and 18\% of cells.

\subsection{Q4: Reuse, sharpness, and computation}
\label{sec:q4}

\textbf{One release, many comparisons, repeated draws.} To assess the main grids' single-draw variability, we drew 50 releases per $\epsilon$ on Beibei at 70\% linkage (first hash mask), each serving all six pairs under Proposition~\ref{prop:dp}. The box covered the counts in 48 of 50 releases at both $\epsilon=1$ and 4, and every interval contained its full-cohort difference. Releases proved 0.9 and 1.0 directions on average, never wrongly. An interval's width had standard deviation at most 0.4\% of its mean, indicating stability in this configuration.

\textbf{What the joint assignment adds.} The same counts also give a simpler valid bound that keeps the item counts (or their boxes) and the known total $|W|$ but drops the constraint that each unlinked user has exactly one outcome: each unit of item $j$ is credited to the most (or least) favorable unlinked users that admit $j$, and a user may be credited for several items. With exact cohort-level counts this item-only bound is nearly as tight as the joint program (on hash cells, 32\% of the per-user width against 29\% on Beibei and 22\% against 20\% on Tmall) and decides the same cells, because each count is small relative to the number of unlinked users that could absorb it. The one-outcome-per-user constraint matters when the information is noisy or fine. On the repeated releases above, the joint intervals are 12\% narrower at $\epsilon=1$ and decide 44 of 300 comparisons (15\%), all EASE against Pop, against none for the item-only bound. At $\epsilon=4$, both decide 50 of 300 comparisons (17\%), again all EASE against Pop. With exact counts for 16 random segments at 90\% linked, the joint program decides ItemKNN against PureSVD in both masks, which the item-only bound first does with 64 segments.

\textbf{Sharpness and computation.} Across 2,000 enumerated small instances, the full LP matched enumeration, the reduced bounds were valid and all 3,024 certified endpoints matched. Beibei's first pass certified 73\% of exact-count cells. The remaining 184 involved Pop and were certified by refinement, moving endpoints by at most 0.0715. With refinement, certification reached 100\% for Tmall exact counts, 99\% for Beibei private counts and 100\% for Tmall private counts. Subsample and segment runs used first-pass bounds (95\% and 69\% certified). Uncertified bounds remain valid outer intervals. All proven directions in Table~\ref{tab:decisions}'s full grids and Fig.~\ref{fig:ladder} clear zero by at least 5.2 in summed contribution $nL$ or $-nU$. The smallest reported margin, 0.044 in a Q4 private release, is confirmed by weak duality using solver multipliers and an interior-point re-solve.

With half of the users unlinked, the reduced network matched the full formulation's endpoints in 0.41\,s instead of 88\,s for 4,000 Beibei users. On one CPU core a cell took 6.1\,s on average on Beibei (first pass) and 49\,s on Tmall with refinement, where the full formulation would have about $5\times10^8$ variables per bound.

\section{Related Work}
\label{sec:related}

\textbf{Evaluation with missing or biased outcomes.} Inverse propensity weighting~\cite{schnabel2016recommendations} and doubly robust estimation~\cite{wang2019doubly} recover point estimates when propensity or outcome models are correct. Our bounds require neither a propensity nor an outcome model and return an identified interval. In information retrieval, incomplete judgments motivated bpref~\cite{buckley2004retrieval}, sampled estimation~\cite{aslam2006statistical}, and bootstrapped NDCG~\cite{frobe2023bootstrapped}, and maximized effectiveness difference~\cite{tan2015med} bounds the gap between two rankings. Carterette et al.~\cite{carterette2006minimal} use the difference in average precision under unknown judgments to guide further judging. Our per-user baseline is the signed one-item case of maximized effectiveness difference. We additionally impose released item counts that couple users.

\textbf{Partial identification and ecological inference.} Bounding a target by the set of distributions consistent with the data is the program of partial identification~\cite{manski2003partial}. Deterministic bounds for ecological inference go back to Duncan and Davis~\cite{duncan1953alternative}, and Cross and Manski~\cite{cross2002regressions} give sharp regions when outcome and covariate are observed in separate data sets. Statistical matching studies the same kind of problem when two data sources share only some variables~\cite{dorazio2006statistical}, and statistical disclosure control bounds the cells of a contingency table given released margins~\cite{dobra2000bounds}. Our per-user baseline is Manski's worst-case bound for missing outcomes~\cite{manski1989anatomy} applied to a ranking contrast. Fan et al.~\cite{fan2014combination} derive sharp treatment-effect bounds under data combination, and Ekeland et al.~\cite{ekeland2010transport} formulate compatibility with incomplete models through optimal transport. Our formulation specializes these marginal-compatibility ideas to finite user-item assignments with known ranking contributions. The contribution is the ranker-evaluation formulation, the reduced network with a sharpness certificate, and the reusable private-count extension. Learning from label proportions~\cite{quadrianto2009estimating} and from private aggregate responses~\cite{javanmard2024learning} instead estimates model parameters from such aggregates.

\textbf{Privacy-preserving measurement.} In advertising, Delaney et al.~\cite{delaney2024dp} formalize differentially private conversion measurement and its attribution and adjacency choices, and Aksu et al.~\cite{aksu2024summary} optimize summary reports under contribution budgets. The W3C Attribution draft specifies aggregated histograms with Laplace noise~\cite{w3c2026attribution}. Confidence intervals from private releases~\cite{karwa2018finite} and private recommender training~\cite{mcsherry2009differentially} address estimation and learning rather than proofs of ranker order. Answering many aggregate queries accurately allows reconstruction of records~\cite{dinur2003revealing}, and our segment experiment observes a small-scale version. Our contribution is a downstream use: turning an item histogram into ranker comparisons with guarantees valid for every comparison drawn from one release.

\section{Discussion}
\label{sec:discussion}

\textbf{Relation to other designs.} When linkage can be taken as random, complete-case confidence intervals are narrower. The aggregate interval remains valid when that assumption fails. Both can be reported: the aggregate one for what holds without a linkage model, the complete-case one for what follows under random linkage. An aggregator that may join slates and outcomes could instead release the metric difference with Laplace noise of scale $2/(n\epsilon)$, but rankers not covered by that release need new queries and privacy budget. An item histogram needs no join with slates and serves any later comparison.

\textbf{Designing the release.} The experiments suggest four design choices. Counts should be restricted to the cohort and the designated outcome (Section~\ref{sec:setting}) and be complete, since a few percent of missing reports removed decisions (Section~\ref{sec:robust}). Under differential privacy, coarse segments were preferable to fine ones. Releasing popular items individually and merging the long tail into a few popularity bins recovered much of the narrowing that a full item histogram loses to noise at $\epsilon=1$.

\textbf{Scope.} Our experiments follow the offline leave-one-out protocol and, in Section~\ref{sec:robust}, its extension to several relevant items per user with counts known to the evaluator. A private multi-item release needs sensitivity $2\bar r$ for a public per-user cap $\bar r$. Neither dataset records consent events, so the linkage patterns are constructed and the decision rates refer to these patterns. A deployment using a discrete Laplace sampler would calibrate the box with that sampler's tail bound, and data-adaptive confidence sets or consistency post-processing~\cite{hay2010boosting} could tighten it.

\section{Conclusion}

Item-level counts can support recommender evaluation after individual outcome links are lost. A transportation formulation bounds the cohort-level performance difference over all consistent assignments, and a reduced network makes these bounds practical for full catalogs. In two e-commerce cohorts, exact counts narrowed every pairwise interval and established many comparisons against Pop. Personalized comparisons were harder, although exact segment counts established both tested Beibei pairs at 90\% linkage. Private counts retained more utility in larger cohorts and coarser releases. One release thus supports later cohort comparisons by proving an order when the counts determine one and quantifying the remaining uncertainty otherwise.

\section*{Acknowledgment}
Claude (Anthropic) assisted with manuscript drafting. ChatGPT (OpenAI) assisted with camera-ready revisions to the abstract and explanatory text in Sections I--VIII, and with scripts that summarize existing experimental outputs.

\bibliographystyle{ieeetr}
\bibliography{references}

\end{document}